\documentclass[11pt]{article}
\usepackage{amssymb,latexsym,amsmath}
\usepackage[dvips]{graphicx}
\usepackage{cite}
\newtheorem{theo}{Theorem}

\newtheorem{lemm}[theo]{Lemma}

\def\ii{\mathrm{i}} % imaginary unit i
\def\ee{\mathrm{e}} % exponential unit e
\newcommand{\myatop}[2]{\genfrac{}{}{0pt}{}{#1}{#2}}
\def\mybox{\hfill$\Box$}%
\begin{document}
\begin{center}
{\Large \bf
The Wigner distribution function for the $\mathfrak{su}(2)$ finite oscillator\\[2mm]
and Dyck paths}\\[5mm]
{\bf Roy Oste, Joris Van der Jeugt} \\[1mm]
Department of Applied Mathematics and Computer Science,
Ghent University,\\
Krijgslaan 281-S9, B-9000 Gent, Belgium\\[1mm]
E-mail: Roy.Oste@UGent.be; Joris.VanderJeugt@UGent.be
\end{center}

\vskip 10mm
\noindent
Short title: $\mathfrak{su}(2)$ finite oscillator and Dyck paths


%\addtolength{\baselineskip}{2mm}
%\addtolength{\abovedisplayskip}{1mm}
%\addtolength{\belowdisplayskip}{1mm}
%\addtolength{\parskip}{1mm}

\begin{abstract}
Recently, a new definition for a Wigner distribution function for a one-dimensional finite quantum system, in which
the position and momentum operators have a finite (multiplicity-free) spectrum, was developed.
This distribution function is defined on discrete phase-space (a finite square grid), and can thus be referred to as
the Wigner matrix. 
In the current paper, we compute this Wigner matrix (or rather, the pre-Wigner matrix, which is related to the
Wigner matrix by a simple matrix multiplication) for the case of the $\mathfrak{su}(2)$ finite oscillator.
The first expression for the matrix elements involves sums over squares of Krawtchouk polynomials, and follows
from standard techniques.
We also manage to present a second solution, where the matrix elements are evaluations of Dyck polynomials. 
These Dyck polynomials are defined in terms of the well known Dyck paths. 
This combinatorial expression of the pre-Wigner matrix elements turns out to be particularly simple.
\end{abstract}

\section{Introduction}

In a previous paper~\cite{VdJ2013}, a new definition of a Wigner distribution function for a finite quantum
system~\cite{Wootters1987,Barker2000,Ozaktas2000}
was developed, in particular for a quantum system in which the position operator $\hat q$ has a finite multiplicity-free
spectrum $\{q_0, q_1, \ldots, q_N\}$ and the momentum operator $\hat p$ also has a finite multiplicity-free
spectrum $\{p_0, p_1, \ldots, p_N\}$. When the system is in a stationary state $|n\rangle$ (i.e.\ the eigenstate
of some Hamiltonian operator $\hat H$; $n=0,1,\ldots,N$), the corresponding Wigner function $W(n;p,q)$ is defined
on a square grid $(p,q) \in \{p_0, p_1, \ldots, p_N\} \times \{q_0, q_1, \ldots, q_N\}$, so it can be considered
as an $(N+1)\times(N+1)$ matrix (the Wigner matrix). 
The assumption in~\cite{VdJ2013} is that --- as in the continuous case~\cite{Tatarskii1983,Hillery1984,Lee1995,Curtright2012}
--- the distribution averages for all ``physical observables'' of the form $p^aq^b$
($a,b=0,\ldots,N$) coincide with their quantum state averages for the corresponding operator form (following
Weyl's association scheme).
This approach led to a procedure to compute Wigner matrices, involving Vandermonde matrices and their inverses.
The Wigner matrices or discrete functions thus defined also satisfy a number of properties similar to those of 
continuous Wigner distribution functions.

One of the simplest examples of a finite quantum system is the so-called $\mathfrak{su}(2)$ finite oscillator model,
introduced by Atakishiyev {\em et al}~\cite{Atak2001,Atak2005}.
For this oscillator, the physical operators (Hamiltonian, position, momentum) are elements of $\mathfrak{su}(2)$, 
acting in a finite (irreducible) $\mathfrak{su}(2)$ representation of dimension $2j+1$ (where $j$ is a nonnegative integer).
The position and momentum operators have as spectrum the set $\{-j,-j+1,\ldots,+j\}$, and the corresponding
discrete wavefunctions are given in terms of Krawtchouk polynomials.
For $j\rightarrow \infty$, these wavefunctions (after rescaling) tend to the common continuous wavefunctions of the harmonic oscillator
in terms of Hermite functions.

In~\cite{VdJ2013}, the discrete Wigner function for the $\mathfrak{su}(2)$ oscillator was computed for some particular
values of $j$, and the matrix plots of these functions were considered. 
This led to quite remarkable observations, in particular when compared to plots of continuous Wigner functions
for the canonical oscillator.

The $\mathfrak{su}(2)$ finite oscillator model is so simple that we cannot be satisfied just with some numerical computations
and plots of the discrete Wigner functions.
Instead, one should be able to give explicit values $W(n;p_k,q_l)$ for the elements of the Wigner matrix (as a
function of $j$, $n$, $k$ and $l$).
This problem is the topic of the current paper, and turns out to lead to interesting mathematics. 
In fact, we have more than one solution for it. 
One approach is based on special functions (hypergeometric series), and follows in a rather straightforward way
from ``basis transformations''.
The second approach is combinatorial. We shall see that the elements of the Wigner matrix
(or, more precisely, the pre-Wigner matrix, to be defined in the following section) coincide with
(the evaluation of) Dyck polynomials. A Dyck polynomial is a multi-variable polynomial ``counting'' all
Dyck paths of a certain type. These Dyck paths themselves are simple combinatorial objects appearing
in many contexts. 

In the following section we recall the definition of the Wigner function for a finite quantum system
and for the $\mathfrak{su}(2)$ finite oscillator.
In section~3 we shall compute the matrix elements of the pre-Wigner matrix using basis transformations,
and obtain an expression in terms of ${}_2F_1$-series or Krawtchouk polynomials.
In section~4, we introduce Dyck paths (which are known) and the corresponding Dyck polynomials (which are new).
We illustrate these combinatorial objects with some examples, which help to understand them.
In section~5 we resume the computation of the pre-Wigner matrix for the $\mathfrak{su}(2)$ finite oscillator,
and see that this is indeed related to Dyck polynomials.

\section{The Wigner function for the $\mathfrak{su}(2)$ finite oscillator}

For a one-dimensional quantum system, let ${\hat H}$, ${\hat q}$ and ${\hat p}$ denote the Hamiltonian, the position and the momentum operator. For a finite quantum system, one assumes that ${\hat H}$, ${\hat q}$ and ${\hat p}$ are self-adjoint elements of some algebra (with a $\star$-operation), and that the unitary representations of this algebra are finite-dimensional.
When these operators moreover satisfy the Hamilton-Lie equations
$[\hat H, \hat q] = -\ii \hat p$, %\qquad 
$[\hat H,\hat p] = \ii \hat q$,
%\label{Hqp}
and when ${\hat H}$ has an equidistant spectrum, the quantum system is referred to as a finite quantum 
oscillator~\cite{Atak2001,Atak2005,JSV2011,JV2012}.

We shall recall some general notation and concepts, introduced in~\cite{VdJ2013}.
Consider a representation space $V$ of dimension $N+1$, and denote the (normalized) eigenvectors of ${\hat H}$ by $|n\rangle$,
with ${\hat H} |n\rangle = E_n |n\rangle$ ($n=0,1,\ldots,N$). $V$ is a Hilbert space, and $\langle n' | n\rangle = \delta_{n',n}$.
These eigenvectors are referred to as the stationary states.
In the basis $|n\rangle$ ($n=0,1,\ldots,N$), the operators ${\hat q}$ and ${\hat p}$ are represented by Hermitian matrices, whose eigenvalues are real and correspond to the finite spectrum of these operators, or the ``possible position and momentum values''.
Let us denote the eigenvalues of ${\hat q}$ by $q_k$ ($k=0,1,\ldots,N$), and the corresponding (orthonormal) eigenvectors by $|q_k\rangle$. We shall assume that all eigenvalues are different and simple (i.e.\ non-degenerate spectra).
So we have the following situation:
\begin{equation}
{\hat q} |q_k\rangle = q_k |q_k\rangle \qquad (k=0,1,\ldots,N).
\label{q-action}
\end{equation}
The expansion of the (orthonormal) ${\hat q}$ eigenstates in the basis $|n\rangle$ ($n=0,1,\ldots,N$) is denoted by
\begin{equation}
|q_k\rangle = \sum_{n=0}^N \phi_n(q_k) |n\rangle \qquad (k=0,1,\ldots,N),
\label{qk}
\end{equation}
and the inverse relation reads
\begin{equation}
|n\rangle = \sum_{k=0}^N \phi_n^\star(q_k) |q_k\rangle \qquad (n=0,1,\ldots,N).
\label{qk-inverse}
\end{equation}
The discrete function $\phi_n(q)$, defined for $q\in\{q_0,q_1,\ldots,q_N\}$, can be interpreted as the position wavefunction when the system is in the $n$th stationary state~\cite{Atak2001}.

Completely similar, we shall assume that the eigenvalues of ${\hat p}$ are given by the mutually distinct values $p_k$ ($k=0,1,\ldots,N$), and denote the corresponding (orthonormal) eigenvectors by $|p_k\rangle$. 
So ${\hat p} |p_k\rangle = p_k |p_k\rangle$ ($k=0,1,\ldots,N$), and we denote
$|p_k\rangle = \sum_{n=0}^N \psi_n(p_k) |n\rangle$ ($k=0,1,\ldots,N$).

In the continuous case, the Wigner function $W_n(p,q)$  is a distribution function
in $(p,q)$-phase space such that the expectation value for a classical phase space function $G(p,q)$ coincides with
the quantum mechanical expectation value of the suitably ordered operator expression $\hat G(\hat p, \hat q)$ for the $n$th stationary state.
In the discrete case, the Wigner function $W(n;p_k,q_l)$ is (for every $n$) a function of the discrete values $(p_k,q_l)$ such that~\cite{VdJ2013}
\begin{equation}
\langle n | {\hat G}({\hat p},{\hat q}) | n\rangle = \sum_{k=0}^N \sum_{l=0}^N W(n;p_k,q_l) G(p_k,q_l),
\label{W1}
\end{equation}
where $G(p,q)$ is supposed to be a polynomial expression in $p$ and $q$, and ${\hat G}({\hat p},{\hat q})$ is the
corresponding operator expression according to Weyl's association.

It will be convenient to represent the Wigner distribution function as an $(N+1)\times(N+1)$ matrix ${\mathbf W}(n)_{0\leq k,l \leq N}$ with matrix elements
\begin{equation}
{\mathbf W}(n)_{k,l} \equiv W(n;p_k,q_l).
\end{equation}

Recall that for a monomial function $G$, denoted by $G_{a,b}(p,q)=p^a q^b$ ($a,b=0,1,\ldots$),
the corresponding operator function would be given by
\begin{equation}
{\hat G}_{a,b}({\hat p},{\hat q}) = \frac{1}{\binom{a+b}{a}} 
 \left. (\lambda \hat p + \mu \hat q)^{a+b} \right|_{\lambda^a\mu^b}.
\label{Gab}
\end{equation}
The last notation stands for taking the coefficient of $\lambda^a\mu^b$ in the expansion of $(\lambda \hat p + \mu \hat q)^{a+b}$.
The technique to construct the discrete Wigner function or matrix was then analyzed in~\cite{VdJ2013}.
First, one should determine a pre-Wigner matrix ${\mathbf Z}(n)_{0\leq a,b\leq N}$ by
\begin{equation}
{\mathbf Z}(n)_{a,b} = \langle n | {\hat G}_{a,b}({\hat p},{\hat q}) |n \rangle \qquad (a,b=0,1,\ldots,N).
\label{defZ}
\end{equation}
Then, the actual Wigner matrix ${\mathbf W}(n)_{0\leq k,l\leq N}$ is given by
\begin{equation}
{\mathbf W}(n) = {\mathbf V}(p_0,p_1\ldots,p_N)^{-T}{\mathbf Z}(n){\mathbf V}(q_0,q_1\ldots,q_N)^{-1}.
\label{main}
\end{equation}
Herein, ${\mathbf V}(x_0,x_1\ldots,x_N)$ stands for the $(N+1)\times(N+1)$ Vandermonde matrix corresponding to the values $x_0,x_1\ldots,x_N$, ${\mathbf A}^{-1}$ for the inverse of a matrix ${\mathbf A}$, and
${\mathbf A}^{-T}$ for the inverse of the transpose of a matrix ${\mathbf A}$.
Since the inverses of Vandermonde matrices are known explicitly~\cite[Lemma~1]{VdJ2013}, 
it is sufficient to compute the pre-Wigner matrices ${\mathbf Z}(n)$.

\vskip 3mm
Let us now turn to the example for which we wish to compute the pre-Wigner matrices ${\mathbf Z}(n)$ explicitly,
the $\mathfrak{su}(2)$ oscillator model~\cite{Atak2001,Atak2005}. 
In terms of the standard $\mathfrak{su}(2)$ basis $J_0, J_+, J_-$ (with commutation relations $[J_0,J_\pm]=\pm J_\pm$, $[J_+,J_-]=2J_0$), and working in the representation space $V=V_j$ labeled by a nonnegative integer of half-integer $j$ ($2j\in{\mathbb Z}_+$),
the Hamiltonian, position and momentum operators are defined by
\begin{equation}
{\hat H}= J_0+j+\frac12,\qquad {\hat q}=\frac12 (J_++J_-), \qquad {\hat p}=\frac{\ii}{2} (J_+-J_-).
\label{Hqp}
\end{equation}
The basis states of $V_j$ in the ``angular momentum'' notation are $|j,m\rangle$ ($m=-j,-j+1,\ldots,+j$), with the well known action
\begin{equation}
J_0 |j,m\rangle = m |j,m\rangle, \qquad J_\pm |j,m\rangle = \sqrt{(j\mp m)(j\pm m+1)} |j,m\pm1\rangle.
\label{Jm-action}
\end{equation}
Thus also the matrices of ${\hat p}$ and ${\hat q}$ in this basis are clear from this action.
Following the earlier notation, we have $N=2j$, and the Hamiltonian eigenstates are denoted by
\begin{equation}
|n\rangle = |j,m\rangle \hbox{ with } n=j+m \qquad (n=0,1,\ldots,N=2j).
\label{n-basis}
\end{equation}
The eigenvalues and eigenvectors of ${\hat q}$ (and ${\hat p}$) have been determined in~\cite{Atak2001}. One has
\[
q_k = p_k = -j+k \qquad (k=0,1,\ldots,N),
\]
with (following the notation of~\eqref{qk}) 
\begin{equation}
\phi_n(q) = \frac{(-1)^n}{2^j} \sqrt{\binom{2j}{n} \binom{2j}{j+q}} K_n(j+q;\frac12,2j).
\label{Kraw1}
\end{equation}
Herein, $K_n$ is the Krawtchouk polynomial~\cite{Koekoek,Ismail,Andrews}:
\begin{equation}
K_n(x;p,N) = {\;}_2F_1 \left( \myatop{-n,-x}{-N} ; \frac{1}{p} \right).
\label{Kraw2}
\end{equation}
So the discrete position wavefunctions are symmetric (i.e.\ with $p=1/2$) Krawtchouk polynomials.
For some plots of these discrete wavefunctions, we refer to~\cite{Atak2001,Atak2005}.
In~\cite{VdJ2013}, we have computed (numerically) some examples of Wigner matrices ${\mathbf W}(n)$
for this $\mathfrak{su}(2)$ case, and given some matrix plots of these discrete Wigner functions.
These plots are interesting, and the shapes of the discrete plots are reminiscent of the shapes
of the continuous plots for the canonical oscillator.

Given the importance of this example, it would be interesting to find general expressions for the
matrix elements of the Wigner matrix ${\mathbf W}(n)$ or of the pre-Wigner matrix ${\mathbf Z}(n)$.
This is the purpose of the current paper.
The first solution is presented in the next section. 
Surprisingly, there is also a combinatorial solution, related to Dyck paths. This will be discussed in sections~4--5.

\section{Computation of the pre-Wigner matrix}

The purpose of this section is the explicit calculation of ${\mathbf Z}(n)_{a,b}$, given
by~\eqref{defZ} and~\eqref{Gab}, for the $\mathfrak{su}(2)$ oscillator model with $\hat p$ and $\hat q$
given by~\eqref{Hqp} with action~\eqref{Jm-action} in the representation space $V_j$ of dimension $2j+1=N+1$.
Note that, using the notation~\eqref{n-basis} for the basis of the representation space, we have
\begin{equation}
J_+ |n\rangle = \sqrt{(n+1)(N-n)} |n+1\rangle,\qquad J_-|n\rangle = \sqrt{n(N+1-n)} |n-1\rangle.
\label{J-action}
\end{equation}
Following~\eqref{Gab} and \eqref{defZ}, we need to compute diagonal entries of powers of $\lambda \hat p + \mu \hat q$.
Note that
\begin{equation}
\lambda \hat p + \mu \hat q = \alpha J_+ + \beta J_-
\end{equation}
where
\begin{equation}
\alpha = \frac12(\mu+\ii \lambda),\qquad \beta=\frac12(\mu-\ii\lambda).
\label{ab}
\end{equation}
Consider $(\alpha \tilde J_+ + \beta \tilde J_-)^r|n\rangle$ for some positive power $r$. 
Due to the simple action~\eqref{J-action}, comparing terms with the same ordered
string of $J_+$'s and $J_-$'s in the expansion of $(\alpha J_+ + \beta J_-)^r$
and of $(J_+ + J_-)^r$ it is easy to see that
\begin{equation}
\langle n+r-2a|(\alpha J_+ + \beta J_-)^r|n\rangle =
\alpha^{r-a}\beta^a \langle n+r-2a| (J_+ + J_-)^r |n\rangle,
\label{XY}
\end{equation}
for $a=0,1,2,\ldots$ (and $0\leq n+r-2a\leq N$).
Since we only need diagonal elements of powers of $\alpha J_+ + \beta J_-$, we can restrict to even powers (the odd powers
having zeros on the diagonal, by~\eqref{XY}).
For even powers, \eqref{XY} yields
\begin{align}
\langle n| (\alpha J_+ + \beta J_-)^{2r} |n\rangle 
 &= \alpha^r \beta^r \langle n| (J_+ + J_-)^{2r} |n\rangle \nonumber\\
 &= \frac{1}{4^r}(\lambda^2+\mu^2)^r \langle n| (J_+ + J_-)^{2r} |n\rangle \nonumber\\
 &= \sum_{i=0}^r \binom{r}{i}\lambda^{2i}\mu^{2r-2i} \langle n| \left(\frac{J_+ + J_-}{2}\right)^{2r} |n\rangle \nonumber\\
 &= \sum_{i=0}^r \binom{r}{i}\lambda^{2i}\mu^{2r-2i} \langle n| {\hat q}^{2r} |n\rangle .
\label{expansion}
\end{align}
So using \eqref{Gab}, \eqref{defZ} and this expansion, one can see that ${\mathbf Z}(n)_{a,b}$ is nonzero only for even values
of the indices $a$ and $b$, and that
\begin{equation}
{\mathbf Z}(n)_{2a,2b}= \frac{\binom{a+b}{a}}{\binom{2a+2b}{2a}}
\langle n| {\hat q}^{2a+2b} |n\rangle.
\label{ZY}
\end{equation}
Hence the computation is reduced to calculating diagonal entries of even powers of $\hat q$ in the $|n\rangle$-basis.
Performing the basis transformation~\eqref{qk-inverse}, this gives (for $r\in\mathbb{Z}_+$): 
\begin{equation*}
\langle n| {\hat q}^{2r} |n\rangle 
 = \sum_{k=0}^N\sum_{l=0}^N \phi_n(q_l)\phi_n^\star(q_k)\; \langle q_l| {\hat q}^{2r} |q_k\rangle,
\end{equation*}
and thus by the action~\eqref{q-action} and orthogonality:
\begin{equation}
\langle n| {\hat q}^{2r} |n\rangle 
 = \sum_{k=0}^N q_k^{2r} |\phi_n(q_k)|^2.
\label{q2r-1}
\end{equation}
Since $\phi_n(q_k)$ is a Krawtchouk polynomial and $q_k=-j+k$, one can use~\eqref{Kraw1} and~\eqref{Kraw2} to find the following 
equivalent expressions:
\begin{align}
\langle n| {\hat q}^{2r} |n\rangle 
 &=\sum_{k=0}^{2j} (-j+k)^{2r} \frac{1}{2^{2j}} \binom{2j}{n}\binom{2j}{k} \left(K_n(k;\frac12;2j)\right)^2 \nonumber\\
 &= \sum_{k=0}^{2j} (-j+k)^{2r} \frac{1}{2^{2j}} \binom{2j}{n}\binom{2j}{k} \left( {\;}_2F_1 \left( \myatop{-n,-k}{-2j} ; 2 \right) \right)^2 \nonumber \\
 &= \sum_{k=0}^{2j} (-j+k)^{2r} \frac{1}{2^{2j}} \binom{2j}{n}\binom{2j}{k} \left( 
 \sum_{i=0}^{\min(n,k)} \frac{\binom{n}{i}\binom{k}{i}}{\binom{2j}{i}} (-2)^i \right)^2.
\label{res1} 
\end{align} 
So this gives us explicit expressions for the matrix elements~\eqref{ZY} of the pre-Wigner matrix ${\mathbf Z}(n)$.

As a simple example, let us consider the pre-Wigner matrix ${\mathbf Z}(0)$ for the ground state, that gives rise to the
ground state Wigner matrix using~\eqref{main}. Using~\eqref{res1} yields: 
${\mathbf Z}(0)_{a,b}$ is nonzero only for even values
of the indices $a$ and $b$ and 
\begin{equation}
{\mathbf Z}(0)_{2a,2b}= \frac{1}{2^{2j}} \frac{\binom{a+b}{a}}{\binom{2a+2b}{2a}} \sum_{k=0}^{2j} \binom{2j}{k} (-j+k)^{2a+2b}.  
\label{res0}
\end{equation}

Eqs.~\eqref{ZY} and~\eqref{res1}, together with the explicit expressions of the Vandermonde matrix inverses, give 
an explicit expression for the Wigner matrix ${\mathbf W}(n)$ for the $\mathfrak{su}(2)$ finite oscillator.
Next to this solution in terms of multiple sum expressions, we can present a second solution in terms of combinatorial quantities.
This is the subject of the following sections.

\section{Dyck paths and polynomials}

A {\em Dyck path $p$ of size $r$} is a lattice path from $(0,0)$ to $(2r,0)$ 
in the integer plane consisting of $r$ up steps of the form $(1,1)$
and $r$ down steps of the form $(1,-1)$ which never passes below the 
$x$-axis~\cite{Goulden1983, Stanley1986, Deutsch1998, Deutsch1999, Mansour2002}. 
An example of a Dyck path of size $5$ is given in figure~1. 
Clearly, one can encode such a Dyck path by a {\em Dyck word} describing the up steps by the letter $u$ and the downs steps by
the letter $d$. So the Dyck path of figure~1 has $uududduudd$ as Dyck word. Formally, a Dyck word of size $r$ is a word consisting of $r$ letters $u$ and $r$ letters $d$ in such a way that (counting from the left) the $u$ count is always greater than or equal
to the $d$ count. 

We will denote by ${\cal D}_r$ the set of all Dyck paths of size $r$ (and by ${\cal D}_0$ the set consisting of the
empty path only). In figure~2, we have listed all elements of ${\cal D}_3$; alternatively, these could be
described by the words~\cite{Deutsch1999}
\begin{equation}
uuuddd, uududd, uuddud, uduudd, ududud.
\label{Dyckword}
\end{equation}
It is well known that the number of Dyck paths of size $r$ is equal to the Catalan number $C_r$.

Various parameters can be defined for Dyck paths~\cite{Deutsch1999,Mansour2002}. 
The {\em height} of a Dyck path $p$ is the largest $i$ for which $p$ touches the line $y=i$. 
In figure~1, the height is $2$. In figure~2, the heights are given by $3,2,2,2$ and $1$ respectively.
Clearly, all elements of ${\cal D}_r$ have height less than or equal to $r$.

We denote by ${\cal D}_{r|h}$ the subset of ${\cal D}_r$ consisting of all elements with height at most $h$.
The notation ${\cal D}_{r|h}$ refers to ``restricting'' to certain elements of ${\cal D}_r$.
So ${\cal D}_{r|r}={\cal D}_{r}$. 
From figure~1, one can see that $|{\cal D}_{3|3}|=5$, $|{\cal D}_{3|2}|=4$ and 
$|{\cal D}_{3|1}|=1$.

We need to consider one further extension (with elements corresponding to a subset of ${\cal D}_{r|h}$).
For a given $r$, and integer values $a,b$ with $0\leq a,b\leq r$, 
let us consider the Dyck paths starting with at least $a$ up steps 
and ending with at least $b$ down steps.
In figure~3 we list the Dyck paths of size $r=5$ with $a=3$ and $b=2$.
We denote by ${\cal D}_{r|h}^{(a,b)}$ the set of all Dyck paths of size $r$ and of height at most $h$, starting
with at least $a$ ups and ending with at least $b$ downs.
So the elements of ${\cal D}_{5|5}^{(3,2)}$ are given in figure~3. 
Note that, clearly, ${\cal D}_{r|h}^{(0,0)}= {\cal D}_{r|h}^{(1,1)}={\cal D}_{r|h}$ 
and ${\cal D}_{r|r}^{(0,0)}= {\cal D}_{r|r} = {\cal D}_r$.

Now we shall introduce a new notion for a Dyck path $p$, namely the {\em weight $w(p)$} of $p$.
An up step of a Dyck path having end points at the integer coordinates $(l,k-1)$ and $(l+1,k)$ is said to be at {\em level $k$}. 
It is convenient to label the integer intervals on the $y$-axis by variables 
$u_1,u_2, \ldots ,u_r$, where the index $k$ in $u_k$
refers to the level. 
In fact, this has been done already in figures 1--3. 
The weight of a path $p$ is the product over all variables $u_i$, in such a way that each up step at level $k$ in $p$ contributes
a factor $u_k$. 
For example, for the path $p$ in figure~1, $w(p)=u_1^2u_2^3$, since there are in total two up steps at level~1 and three up steps at level~2.
The weights of the paths in figure~2 are, respectively:
\begin{equation}
u_1u_2u_3,\quad u_1u_2^2,\quad u_1^2u_2,\quad u_1^2u_2,\quad u_1^3.
\label{Dyckweight}
\end{equation}
The weights of the paths in figure~3 are
\begin{align}
& u_1u_2u_3u_4u_5,\quad  u_1u_2u_3u_4^2,\quad  u_1u_2u_3^2u_4,\quad  u_1u_2^2u_3u_4,\quad  u_1u_2u_3^2u_4,\nonumber\\
& u_1u_2u_3^3,\quad u_1u_2^2u_3^2,\quad u_1u_2^2u_3^2,\quad u_1u_2^3u_3,\quad u_1^2u_2^2u_3.
\end{align}
Clearly, each Dyck word as in~\eqref{Dyckword} corresponds to a weight as in~\eqref{Dyckweight}.
In principle, one could extend the notion of weight by introducing next to the ``up variables'' $u_1,u_2,\ldots$ 
also ``down variables'' $d_1,d_2,\ldots$. Then each Dyck word would correspond to a monomial in the variables $u_i$ and in
the variables $d_i$. 
This extension would not give extra information, however, since ``what goes up must come down''.
For this reason, we shall not consider this extension here and work with ``up variables'' $u_i$ only.

Finally, we define the {\em Dyck polynomial} $P_{r|h}^{(a,b)}$ as the sum of the weights of all elements of ${\cal D}_{r|h}^{(a,b)}$:
\begin{equation}
P_{r|h}^{(a,b)} \equiv P_{r|h}^{(a,b)}(u) \equiv P_{r|h}^{(a,b)} (u_1,\ldots,u_h) = \sum_{p \in {\cal D}_{r|h}^{(a,b)} } w(p).
\label{Dpol}
\end{equation}
For convenience, we shall use a simpler notation for those polynomials corresponding to non-restricted Dyck paths (i.e.\ with
no restriction on height and/or on the start and end):
\begin{align}
& P_r^{(a,b)} \equiv P_{r|r}^{(a,b)},  \\
& P_r  \equiv P_{r}^{(0,0)}.
\end{align}
From the previous examples, one can see:
\begin{align*}
& P_3 = P_{3|3}^{(0,0)} = u_1u_2u_3 + u_1u_2^2+ 2 u_1^2u_2 + u_1^3,\\
& P_{3|2}^{(0,0)}  = u_1u_2^2+ 2 u_1^2u_2 + u_1^3,\\
& P_{5}^{(3,2)} = P_{5|5}^{(3,2)} = u_1u_2u_3(u_4u_5 + u_4^2 + 2u_3u_4 + u_2u_4 + u_3^2 + 2u_2u_3 + u_2^2 + u_1u_2).
\end{align*}

Let us list some properties of these polynomials.
First of all, the elements of ${\cal D}_{r|h}$ are those of ${\cal D}_{r|r}={\cal D}_r$ of height at most $h$. 
The weights of the elements of ${\cal D}_{r|r}\setminus {\cal D}_{r|h}$ have factors $u_{h+1}, u_{h+2},\ldots$.
So by putting $u_{h+1}=\ldots=u_r=0$ in $P_{r|r}^{(0,0)}$, one obtains $P_{r|h}^{(0,0)}$:
\begin{equation}
P_{r|h}^{(0,0)} = P_{r|r}^{(0,0)} (u_1,u_2, \ldots, u_h, 0,\ldots,0) = P_r(u_1,u_2,\ldots,u_h,\ldots,0).
\end{equation}
Clearly, this is also valid for
\begin{equation}
P_{r|h}^{(a,b)} = P_{r}^{(a,b)} (u_1,u_2, \ldots, u_h, 0,\ldots,0).
\end{equation}
For this reason, it will be sufficient to study $P_{r}^{(a,b)}$, and thus work with Dyck paths of size $r$ that are not restricted
in height.

Secondly, we can give a recurrence relation for these polynomials.
Consider first the standard Dyck paths of ${\cal D}_{r}$, with $|{\cal D}_{r}|=C_r$, the Catalan number.
These Catalan numbers satisfy
\begin{equation}
C_{r+1}=\sum_{i=0}^r C_i \cdot C_{r-i}.
\end{equation}
One proof of this recurrence is as follows: the Dyck paths of size $r+1$ are obtained by ``raising'' a Dyck path of size $i$ by one level and concatenating with a Dyck path of size $r-i$. (In terms of the Dyck word, raising means adding one letter $u$ in front and one letter $d$ at the end.) 
This same proof gives rise to:
\begin{equation}
P_{r+1} = \sum_{i=0}^r u_1 P_{i}(u_2,u_3,\ldots,u_{i+1}) \cdot P_{r-i}(u_1,u_2,\ldots,u_{r-i}).
\label{recur1}
\end{equation}
So in this relation the indices of the variables in $P_{i}$ are raised by 1.
The first few polynomials are given by
\begin{align*}
&P_0=1, \quad P_1=u_1\quad, P_2= u_1^2+u_1u_2,\quad P_3= {u_{{1}}}^{3}+2\,{u_{{1}}}^{2}u_{{2}}+u_{{1}}{u_{{2}}}^{2}+u_{{1}}u_{{2}}u_{{3}}, \\
&P_4= {u_{{1}}}^{4}+3\,{u_{{1}}}^{3}u_{{2}}+3\,{u_{{1}}}^{2}{u_{{2}}}^{2}+2
\,{u_{{1}}}^{2}u_{{2}}u_{{3}}+u_{{1}}{u_{{2}}}^{3}+2\,u_{{1}}{u_{{2}}}
^{2}u_{{3}}+u_{{1}}u_{{2}}{u_{{3}}}^{2}+u_{{1}}u_{{2}}u_{{3}}u_{{4}},\\
&P_5= 2{u_{{1}}}^{2}u_{{2}}u_{{3}}u_{{4}}+2u_{{1}}{u_{{2}}}^{2}u_{{3}}u_
{{4}}+2u_{{1}}u_{{2}}{u_{{3}}}^{2}u_{{4}}+u_{{1}}u_{{2}}u_{{3}}{u_{{
4}}}^{2}+4{u_{{1}}}^{4}u_{{2}}+6{u_{{1}}}^{3}{u_{{2}}}^{2}+4{u_{
{1}}}^{2}{u_{{2}}}^{3}  +u_{{1}}{u_{{2}}}^{4}\\
&+{u_{{1}}}^{5}+3\,{u_{{1}}}
^{3}u_{{2}}u_{{3}}+6\,{u_{{1}}}^{2}{u_{{2}}}^{2}u_{{3}}+2\,{u_{{1}}}^{
2}u_{{2}}{u_{{3}}}^{2}+3\,u_{{1}}{u_{{2}}}^{3}u_{{3}}+3\,u_{{1}}{u_{{2
}}}^{2}{u_{{3}}}^{2}+u_{{1}}u_{{2}}{u_{{3}}}^{3}+u_{{1}}u_{{2}}u_{{3}}
u_{{4}}u_{{5}}.
\end{align*}

In a similar way, one obtains a recurrence relation for $P_{r}^{(a,b)}$.
First of all, note that obviously
\begin{align}
%P_{r,r}^{(\ell,k)}=0 \hbox{ for } \ell>r \hbox{ or } k>r.
& P_{r}^{(a,b)}=0 \hbox{ for }r<0 \hbox{ or } a<0 \hbox{ or }b<0, \label{zero1} \\
& P_r^{(a,b)}=0 \hbox{ for }a>r \hbox{ or } b>r. \label{zero2}
\end{align}
Furthermore, since all Dyck paths start with at least one up step and end with at least one down step, one has
\begin{equation}
P_{r}^{(0,0)} = P_{r}^{(1,0)} = P_{r}^{(0,1)} = P_{r}^{(1,1)}=P_r \qquad (r>0).
\label{boundary}
\end{equation}
The previous concatenating procedure then gives rise to, for $(a\geq 2,b\geq 1)$ or $(a\geq 1, b\geq 2)$:
\begin{align}
P_{r+1}^{(a,b)} & = \sum_{i=0}^{r-1} u_1 P_{i}^{(a-1,1)}(u_2,u_3,\ldots,u_{i+1}) \cdot 
P_{r-i}^{(1,b)}(u_1,u_2,\ldots,u_{r-i})\nonumber \\
&  + u_1 P_{r}^{(a-1,b-1)}(u_2,u_3,\ldots,u_{r+1}).
\label{recur2}
\end{align}
Note that the sum over $i$ in~\eqref{recur2} runs in fact from $a-1$ to $r-b$ (for $i$ from 0 to $a-2$, $P_i^{(a-1,1)}=0$
due to~\eqref{zero2}; for $i$ from $r-b+1$ to $r-1$, $P_{r-i}^{(1,b)}=0$ for the same reason).
The recurrence relations~\eqref{recur1} and~\eqref{recur2}, together with the boundary conditions~\eqref{zero1},
\eqref{zero2} and \eqref{boundary}, can easily be used to compute the Dyck polynomials $P_r^{(a,b)}$
by means of a computer algebra package.

We close this section with another relation between the Dyck polynomials, which will be useful in the following section.
\begin{lemm}
For $2\leq a\leq r$ and $0\leq b\leq r$, one has
\begin{equation}
P^{(a,b)}_{r+2}- P^{(a+2,b)}_{r+2}  = u_{a-1} u_a P^{(a-2,b)}_{r}  + (u_a + u_{a+1}) P^{(a,b)}_{r+1} .
\label{P-rec}
\end{equation}
\end{lemm}
\noindent  {\bf Proof.}
Consider first the left hand side (lhs) of~\eqref{P-rec}.
As obviously $\mathcal{D}^{(a+2,b)}_{r+2} \subset\mathcal{D}^{(a,b)}_{r+2} $, the lhs is precisely the sum of the 
weights of all elements of $\mathcal{D}^{(a,b)}_{r+2} \setminus\mathcal{D}^{(a+2,b)}_{r+2}$.
Hence, this consists of the weights of all Dyck paths of size $r+2$ that end with at least $b$ down steps and start 
with at least $a$ but at most $a+1$ up steps.

Now it suffices to note that the sum of the weights of the elements of $\mathcal{D}^{(a,b)}_{r+2} \setminus\mathcal{D}^{(a+2,b)}_{r+2}$ can also be computed in another way.
A path in $\mathcal{D}^{(a,b)}_{r+2} \setminus\mathcal{D}^{(a+2,b)}_{r+2}$ has to start with $a$ up steps. If the next step is also an up step we have a path starting with $a+1$ up steps, which necessitates the subsequent step to be a down step. 
This leaves --- removing the up step starting at $(a,a)$ and the down step starting at $(a+1,a+1)$, and putting
the loose ends together --- a remaining path of size $r+1$ starting with at least $a$ up steps and ending with at least $b$ steps down.
The sum of the weights of the corresponding paths in $\mathcal{D}^{(a,b)}_{r+2} \setminus\mathcal{D}^{(a+2,b)}_{r+2}$ is then exactly equal to
\[
 u_{a+1} P^{(a,b)}_{r+1} .
\]
If we have a path with $a$ up steps followed by a down step, the subsequent step can be either up or down. If this last step is up, this leaves a remaining path of size $r+1$ starting at with at least $a$ ups and ending with at least $b$ downs. The sum of the weights of all paths of this type is 
\[
u_a P^{(a,b)}_{r+1}.
\]
In a similar way, the sum of the weights of all the remaining paths is equal to
\[
u_{a-1} u_a P^{(a-2,b)}_{r},
\]
which completes the proof.
\mybox

Note that the relation~\eqref{P-rec} remains valid for $0\leq a\leq r+1$ and $0\leq b \leq r+1$, 
provided one takes into
account~\eqref{zero1}-\eqref{zero2} and $u_{-1}=u_0=0$.

\section{Combinatorial expression for the pre-Wigner matrix}

Let us now turn to a combinatorial computation of ${\mathbf Z}(n)_{a,b}$ for the $\mathfrak{su}(2)$ oscillator model 
in the representation space $V_j$ of dimension $2j+1=N+1$.
For this purpose, it will be helpful to generalize the operators $J_+$ and $J_-$ in~\eqref{J-action} and to introduce
new operators denoted by a tilde:
\begin{equation}
\tilde J_+ |n\rangle = \sqrt{u_{n+1}} |n+1\rangle,\qquad \tilde J_-|n\rangle = \sqrt{u_n } |n-1\rangle,
\label{Jtilde-action}
\end{equation}
where $u_i$ are variables with boundaries $u_0=u_N=0$. 
So under the substitution
\begin{equation}
u_i \rightarrow i(N+1-i)  \qquad(i=1,\ldots,N=2j)
\label{subs}
\end{equation}
the tilde-operators become the genuine operators that we need to use.
We can think of all these operators as $(N+1)\times(N+1)$-matrices relative to
the ordered basis $|0\rangle, |1\rangle, \ldots |N\rangle$ of the representation space $V_j$.
We now consider the extension of the other operators, also denoted by a tilde.
Thus, as in~\eqref{Hqp}, $\tilde q = \frac12(\tilde J_++\tilde J_-)$ and $\tilde p= \frac{\ii}{2}(\tilde J_+ - \tilde J_-)$,  
\begin{equation}
\tilde X = \lambda \tilde p + \mu \tilde q,\qquad
{\tilde G}_{a,b}({\tilde p},{\tilde q}) = \frac{1}{\binom{a+b}{a}} 
 \left. {\tilde X}^{a+b} \right|_{\lambda^a\mu^b},
\label{Xtilde}
\end{equation}
and following~\eqref{Gab} and \eqref{defZ}
\begin{equation}
\tilde {\mathbf Z}(n)_{a,b} = \langle n | {\tilde G}_{a,b}({\tilde p},{\tilde q}) |n \rangle \qquad (a,b=0,1,\ldots,N).
\label{Ztilde}
\end{equation}
Just as in section~3, we need to compute diagonal entries of powers of $\tilde X$, where
\begin{equation}
\tilde X = \lambda \tilde p + \mu \tilde q = \alpha \tilde J_+ + \beta \tilde J_-
\end{equation}
with $\alpha$ and $\beta$ given in~\eqref{ab}.
The same argument as in~\eqref{XY} leads to
\begin{equation}
\langle n+r-2a|(\alpha \tilde J_+ + \beta \tilde J_-)^r|n\rangle =
\alpha^{r-a}\beta^a \langle n+r-2a| (\tilde J_+ + \tilde J_-)^r |n\rangle,
\label{XYtilde}
\end{equation}
for $a=0,1,2,\ldots$ (and $0\leq n+r-2a\leq N$).
The diagonal elements of odd powers of $\tilde X$ are zero, and for the diagonal elements of even powers we find,
as in~\eqref{expansion}:
\begin{align}
\langle n| {\tilde X}^{2r} |n\rangle &= \alpha^r \beta^r \langle n| (\tilde J_+ + \tilde J_-)^{2r} |n\rangle \nonumber\\
&= \frac{1}{4^r}(\lambda^2+\mu^2)^r \langle n| (\tilde J_+ + \tilde J_-)^{2r} |n\rangle \nonumber\\
&= \frac{1}{4^r} \sum_{i=0}^r \binom{r}{i}\lambda^{2i}\mu^{2r-2i} \langle n| (\tilde J_+ + \tilde J_-)^{2r} |n\rangle.
\label{expansion2}
\end{align}
By using \eqref{Xtilde} and \eqref{Ztilde}, one can see that $\tilde {\mathbf Z}(n)_{a,b}$ is nonzero only for even values
of the indices $a$ and $b$, and that
\begin{equation}
\tilde {\mathbf Z}(n)_{2a,2b}= \frac{1}{4^{a+b}} \frac{\binom{a+b}{a}}{\binom{2a+2b}{2a}}
\langle n| (\tilde J_+ + \tilde J_-)^{2a+2b} |n\rangle.
\label{ZYtilde}
\end{equation}
In this case, the problem is reduced to computing diagonal elements of even powers of the tridiagonal matrix
\begin{equation}
Y \equiv \tilde J_+ + \tilde J_- =
\left(\begin{array}{cccccc}
0 & \sqrt{u_1} & 0 & & &\\
\sqrt{u_1} & 0 & \sqrt{u_2} & & &\\
0 & \sqrt{u_2} & 0 & & &\\
 & &\ddots & \ddots &\ddots &\\
 & & & & 0 & \sqrt{u_N} \\
 & & & & \sqrt{u_N} & 0
\end{array}\right).
\label{Y}
\end{equation}
Our main result is that diagonal elements of (even) powers of $Y$ are precisely given by Dyck polynomials:
\begin{equation}
\langle n | Y^{2r} |n\rangle = \frac{P_{r+n|N}^{(n,n)}}{u_1 u_2\cdots u_n}.
\label{mainY}
\end{equation}
This follows from the following theorem, which makes a connection between all matrix elements of even
powers of the matrix $Y$ and Dyck polynomials.
\begin{theo}
\label{theoY}
Let $Y$ be the matrix~\eqref{Y} and $r$ a positive integer. For $0\leq a,b\leq N$ and $a+b$ even, we have
\begin{equation}
\langle a | Y^{2r} |b\rangle = \left(Y^{2r}\right)_{a,b} =  \frac{P_{r+(a+b)/2 | N}^{(a,b)}}{\sqrt{u_1\cdots u_a}\sqrt{u_1\cdots u_b}}
\label{eqtheoY}
\end{equation}
whereas for $a+b$ odd we have $\langle a | Y^{2r} |b\rangle = (Y^{2r})_{a,b} =0$.
\end{theo}
Taking all material together, we arrive at the following expression for the matrix elements of
the pre-Wigner matrix:
\begin{equation}
{\mathbf Z}(n)_{2a,2b}= \frac{1}{4^{a+b}} \frac{\binom{a+b}{a}}{\binom{2a+2b}{2a}}
\left. \frac{P_{a+b+n|N}^{(n,n)}(u)}{u_1 u_2\cdots u_n} \right|_{u_i \rightarrow i(N+1-i)},
\label{final}
\end{equation}
using the substitution~\eqref{subs}.

\vskip 2mm
\noindent {\bf Proof of Theorem~\ref{theoY}.} 
We shall prove~\eqref{eqtheoY} by induction on $r$. Let us first consider the case $r=1$. 
The matrix $Y^2$ has a banded form:
\begin{equation}
Y^2=\left(\begin{array}{ccccccc}
u_1 & 0 & \sqrt{u_1u_2} &   &&& \\
0 & u_1 + u_2 & 0 & \sqrt{u_2u_3} & && \\
\sqrt{u_1u_2} & 0  & u_2 + u_3 & 0 & \sqrt{u_3u_4}  && \\
 & \sqrt{u_2u_3} & 0  & u_3 + u_4 & 0 &\ddots &   \\
 & & \ddots & \ddots & \ddots &\ddots & \sqrt{u_{N-1}u_N} \\
 & & &\ddots&0&u_{N-1}+u_N & 0 \\
 & & & & \sqrt{u_{N-1}u_N} & 0 & u_N
\end{array}\right).
\label{Y2}
\end{equation}
These are the matrix elements in the left hand side of~\eqref{eqtheoY} for $r=1$. 
In the right hand side of~\eqref{eqtheoY}, we should consider the polynomials $P_{1+(a+b)/2 | N}^{(a,b)}$, where
$a+b$ is even. By~\eqref{zero2}, this polynomial is zero if $|a-b|>2$. 
When $\lvert a-b \rvert \leq 2$ we distinguish between the possible cases. 
For $b=a\geq 1$ there are two possible Dyck paths of size $a+1$ starting with at least $a$ up steps and ending with at least $a$ down steps. The weights of these paths are respectively $u_1 \cdots u_{a-1} u_a^2$ and $u_1 \cdots u_au_{a+1}$.
Hence, we find
\[
\frac{P^{(a,a)}_{a+1|N}}{u_1\dotsm u_a} = u_a + u_{a+1}\quad (1\leq a \leq N-1)\quad \hbox{ and }\quad
\frac{P^{(N,N)}_{N+1|N}}{u_1\dotsm u_N} = u_N,
\]
so these match with the entries in~\eqref{Y2}.
For $b=a+2\geq 2$, there is only one possible Dyck path of size $a+2$ starting with at least $a$ up steps and ending with at least $a+2$ down steps, which has associated weight $u_1\cdots u_{a+2}$.
This leads to 
\[
 \frac{P^{(a,a+2)}_{a+2}}{{u_1\cdots u_{a}}\sqrt{u_{a+1}u_{a+2}}}= \sqrt{u_{a+1}u_{a+2}},
\]
matching with the corresponding entries in~\eqref{Y2}. Finally, the case $b=a-2$ is handled similarly.

Now we can use the induction hypothesis, and assume that~\eqref{eqtheoY} holds for $r\geq 1$. Next, we prove it then
holds for $r+1$. The general matrix elements of $Y^{2r+2}$ are computed as follows:
\begin{align*}
& \langle a \,\vert \, Y^{2r+2} \,|\, b \rangle  = \langle a \,\vert \, Y^2 \, Y^{2r} \,|\, b \rangle \\
&= \sum_{n=0}^N \langle a \,\vert \, Y^2 \,|\, n \rangle \langle n \,\vert \, Y^{2r} \,|\, b \rangle  \\
&= \langle a \,\vert \, Y^2 \,|\, a-2 \rangle \langle a-2 \,\vert \, Y^{2r} \,|\, b \rangle +  \langle a \,\vert \, Y^2 \,|\, a \rangle \langle a \,\vert \, Y^{2r} \,|\, b \rangle  + \langle a \,\vert \, Y^2 \,|\, a+2 \rangle \langle a+2 \,\vert \, Y^{2r} \,|\, b \rangle \\
&=  \frac{  \sqrt{u_{a-1} u_a}\, P^{(a-2,b)}_{r+({a+b})/{2}-1|N} }{\sqrt{u_1\dotsm u_{a-2}}\sqrt{u_1\dotsm u_b}} +
 \frac{ (u_a+u_{a+1}) \, P^{(a,b)}_{r+({a+b})/{2}|N} }{\sqrt{u_1\dotsm u_{a}}\sqrt{u_1\dotsm u_b}} +
 \frac{  \sqrt{u_{a+1} u_{a+2}}\, P^{(a+2,b)}_{r+({a+b})/{2}+1|N} }{\sqrt{u_1\dotsm u_{a+2}}\sqrt{u_1\dotsm u_b}}\\
&= \Big( {u_{a-1} u_a}\, P^{(a-2,b)}_{r+({a+b})/{2}-1|N} + (u_a+u_{a+1}) \, P^{(a,b)}_{r+({a+b})/{2}|N}   +
 P^{(a+2,b)}_{r+({a+b})/{2}+1|N} \Big) /{\sqrt{u_1\dotsm u_{a}}\sqrt{u_1\dotsm u_b}}.
\end{align*}
Using~\eqref{P-rec} (which is obviously also valid after restriction to $N$ variables), this gives
\[
\langle a \,\vert \, Y^{2r+2} \,|\, b \rangle  =
 \frac{P^{(a,b)}_{r+({a+b})/{2}+1|N}}{\sqrt{u_1\dotsm u_a}\sqrt{u_1\dotsm u_b}}.
\]
The ``boundary'' matrix elements of $Y^{2r+2}$ are computed similarly. 
\mybox

\section{Conclusions and remarks}

In a previous paper~\cite{VdJ2013}, the notion of Wigner function for a finite quantum system was approached in a new way.
It was illustrated by the (numerical) computation of the discrete Wigner function for the so-called $\mathfrak{su}(2)$ finite oscillator.
Plots of these discrete Wigner functions (over discrete phase-space) led to appealing pictures and interesting observations~\cite{VdJ2013}.

The $\mathfrak{su}(2)$ finite oscillator model is simple, and our goal was to compute the discrete Wigner function explicitly
for this model. This goal has been achieved in this paper, by computing the entries of the pre-Wigner matrix explicitly.
The first solution is given by~\eqref{ZY} and~\eqref{res1}, in terms of special functions (or multiple sums). 
The second solution is given by~\eqref{final}, in terms of the newly introduced Dyck polynomials.

Note that the second approach transcends in fact the $\mathfrak{su}(2)$ model.
Indeed, also for other finite oscillator models in which the $\hat q$ operator has a shape of the form~\eqref{Y}
in the $|n\rangle$ basis of stationary states (for example~\cite{JSV2011} or~\cite{JV2012}), the solution is given
in terms of Dyck polynomials.

A natural question that arises is whether the discrete Wigner function $W(n;p_k,q_l)$ has a continuum limit.
For this purpose, recall that the discrete position wavefunction $\phi_n(q)$ with $q=-j,-j+1,\ldots,+j$, as given in~\eqref{Kraw1}, satisfies
the limit relation~\cite{Atak2005}
\begin{equation}
\lim_{j\rightarrow\infty} j^{1/4} \phi_n(j^{1/2}x) = \frac{1}{2^{n/2}\sqrt{n!}\pi^{1/4}} H_n(x) \ee^{-x^2/2} \equiv \psi_n(x),
\end{equation}
where $H_n$ is the Hermite polynomial and $\psi_n(x)$ is the normalized position wavefunction for the canonical oscillator.
In other words, after a proper rescaling of the finite discrete spectrum, the discrete wavefunctions tend to the continuous wavefunctions
of the canonical oscillator when the representation parameter $j$ tends to infinity.
In a similar fashion, one should consider the discrete Wigner function $W(n;p,q)$, with $p$ and $q$ in $\{-j,-j+1,\ldots,+j\}$.
Using the marginal~\cite[eq.~(30)]{VdJ2013}
\[
\sum_{p=-j}^{+j} W(n;p,q) = |\phi_n(q)|^2, \qquad (q=-j,-j+1,\ldots,+j),
\]
and making the replacement $p=j^{1/2}y$, $q=j^{1/2}x$, one finds 
\[
\int {\cal W}(n;y,x) dy = |\psi_n(x)|^2,
\]
where 
\begin{equation}
{\cal W}(n;y,x) = \lim_{j\rightarrow\infty} j\, W(n;j^{1/2}y, j^{1/2}x).
\label{limit}
\end{equation}
Similarly, using the marginal~\cite[eq.~(31)]{VdJ2013} one obtains:
\[
\int {\cal W}(n;y,x) dx = |\psi_n(y)|^2.
\]
So the limit to consider is $\lim_{j\rightarrow\infty} j\, W(n;j^{1/2}y, j^{1/2}x)$, and on the basis of the above marginals one can expect 
that this limit will be the Wigner function for the canonical oscillator, i.e.
\[
W_n(y,x) = \frac{(-1)^n}{\pi} \ee^{-x^2-y^2} L_n(2x^2+2y^2),
\]
where $L_n$ is the Laguerre polynomial.
This is also confirmed by our plots based on numerical values of the discrete Wigner function, see~\cite{VdJ2013}.
Of course, the marginals alone do not uniquely fix the Wigner functions, so the above argument is not a proof that ${\cal W}(n;y,x)=W_n(y,x)$.
Note that even with the current explicit expressions for $W(n;p_k,q_l)$ obtained in this paper, the computation of the limit~\eqref{limit} 
is still not feasable.
The reason is that one should have some `functional' expression like~\eqref{Kraw1} for $W(n;p,q)$ with $p,q\in\{-j,-j+1,\ldots, +j\}$, before one can make the replacements $p=j^{1/2}y$, $q=j^{1/2}x$.
In our formulas, this is obstructed by the Vandermonde matrix inverses in~\eqref{main}, for which there is an expression as matrix elements but not as functions of $p$ and $q$.

Let us conclude with some remarks that are of mathematical nature.
First of all, note that others have also considered polynomials associated with the set of Dyck paths ${\cal D}_r$.
A particular interesting polynomial --  let us denote it here by $Q_r$ -- was introduced and studied in~\cite{MansourSun} (in a more general setting).
Herein, a {\em $u$-segment} of a path $p$ in ${\cal D}_r$ is defined as a maximal sequence of consecutive up-steps in $p$.
Let $\alpha_i(p)$ be the number of $u$-segments of length~$i$ in~$p$, and
\begin{equation}
Q_r = \sum_{p\in {\cal D}_r} \prod_{i\geq 1} t_i^{\alpha_i(p)}.
\label{Qr}
\end{equation}
For example, for $r=3$, consider the five Dyck paths given in Figure~2. The first path has one $u$-segment of length~3, the second path has one $u$-segment of length~2 and one of length~1, etc. So the five terms in~\eqref{Qr} are, respectively, $t_3$, $t_1t_2$, $t_1t_2$, $t_1t_2$ and $t_1^3$, or:
\[ 
Q_3= t_1^3+3t_1t_2+t_3.
\]
Clearly, the polynomials $Q_r$ are not the same as our Dyck polynomials $P_r$. 
Both are in a way generating functions, but counting quite different statistics: in $P_r$ the {\em height} of each up-step is the crucial
characteristic, whereas in $Q_r$ the length of consecutive up-steps is determinative.
So the two polynomials are very different in nature, and we cannot expect them to be related.
Note that for $Q_r$ there is an expression in terms of partial Bell polynomials~\cite{MansourSun}.
For $P_r$ there is -- as shown in this paper -- a relation with powers of a particular tridiagonal matrix $Y$.

Secondly, for the standard Dyck polynomials introduced in section~4, one can consider the generating function
\[
G(t;u_1,u_2,\ldots)\equiv \sum_{r=0}^\infty P_r t^r = \sum_{r=0}^\infty P_r(u_1,u_2,\ldots,u_r) t^r.
\]
Multiplying the recurrence relation~\eqref{recur1} by $t^{r+1}$ and summing over all $r\geq 0$ then gives:
\begin{align*}
G(t;u_1,u_2,\ldots )-1 &= \sum_{r=0}^\infty \sum_{i=0}^r u_1 P_i(u_2,u_3,\ldots) P_{r-i}(u_1,u_2,\ldots) t^{r+1} \\
&= t u_1 \left(\sum_{i=0}^\infty P_i(u_2,u_3,\ldots) t^i\right) \left(\sum_{r=i}^\infty P_{r-i}(u_1,u_2,\ldots) t^{r-i}\right) \\
&= t u_1 G(t;u_2,u_3,\ldots) G(t;u_1,u_2,\ldots).
\end{align*}
So
\[
G(t;u_1,u_2,\ldots) = \frac{1}{1-t u_1 G(t;u_2,u_3,\ldots)},
\]
and repeated use of this leads to
\begin{equation}
G(t;u_1,u_2,\ldots) = \frac{1}{ \displaystyle 1- \frac{t u_1}{\displaystyle 1- \frac{t u_2}{\displaystyle 1-\frac{t u_3}{\displaystyle 1-\cdots}}}}.
\end{equation}
Clearly, for Dyck paths restricted to height $N$ and the corresponding Dyck polynomials restricted to $N$ variables,
the generating function
\[
G(t;u_1,u_2,\ldots,u_N)=\sum_{r=0}^N P_r t^r + {\cal O}(t^{N+1})
\]
becomes
\begin{equation}
G(t;u_1,u_2,\ldots,u_N) = \frac{1}{ \displaystyle 1- \frac{t u_1}{\displaystyle 1- \frac{t u_2}{\displaystyle 1-\frac{\ddots}{\displaystyle 1-t u_N}}}}.
\end{equation}

As a third remark, note that in this context the symmetric matrix $Y$ in~\eqref{Y} appears naturally.
But of course, one can write $Y'=DYD^{-1}$, with $D=\hbox{diag}(1,\sqrt{u_1},\sqrt{u_1 u_2}, \sqrt{u_1 u_2 u_3}, \ldots)$ and
\begin{equation}
Y'= 
\left(\begin{array}{cccccc}
0 & 1 & 0 & & &\\
u_1 & 0 & 1 & & &\\
0 & u_2 & 0 & & &\\
 & &\ddots & \ddots &\ddots &\\
 & & & & 0 & 1 \\
 & & & & u_N & 0
\end{array}\right).
\label{Y'}
\end{equation}
In this way, one gets rid of the square roots, and for $a+b$ even it follows from~\eqref{eqtheoY} that the matrix elements
of even powers of $Y'$ are essentially Dyck polynomials:
\[
(Y')^{2r}_{a,b} = \frac{P^{(a,b)}_{r+(a+b)/2 |N}}{u_1\cdots u_b}.
\]

Examining the shape of $Y'$ leads to an obvious extension to more general tridiagonal matrices, namely
\begin{equation}
M= 
\left(\begin{array}{cccccc}
h_0 & 1 & 0 & & &\\
u_1 & h_1 & 1 & & &\\
0 & u_2 & h_2 & & &\\
 & &\ddots & \ddots &\ddots &\\
 & & & & h_{N-1} & 1 \\
 & & & & u_N & h_N
\end{array}\right).
\label{M}
\end{equation}
This is related to extensions of Dyck paths to Motzkin paths~\cite{Donaghey}, where apart from up and down steps also
horizontal steps are allowed. One can define the corresponding Motzkin polynomials in a similar way as in~\eqref{Dpol},
and then identify matrix elements of powers of $M$ to specific Motzkin polynomials. This study will be developed elsewhere.

\newpage
\begin{figure}[th]
\[
\includegraphics[scale=0.8]{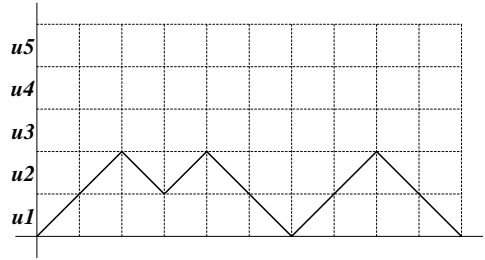}
\]
\caption{Example of a Dyck path of size $r=5$.}
\label{fig1}
\end{figure}

\begin{figure}[h]
\[
\begin{tabular}{ccc}
\includegraphics[scale=0.65]{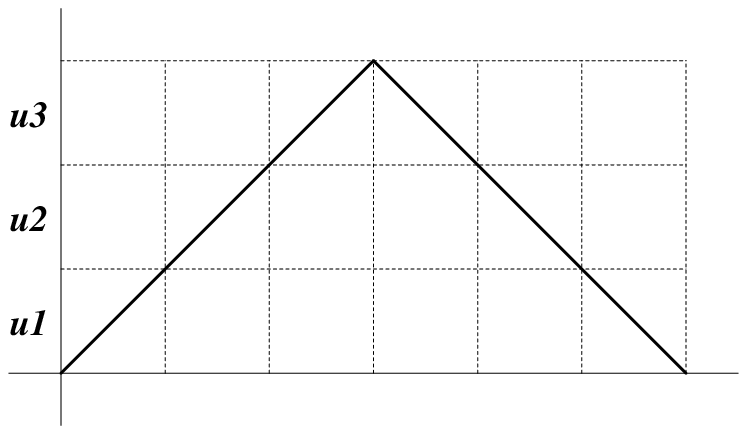} &  \includegraphics[scale=0.65]{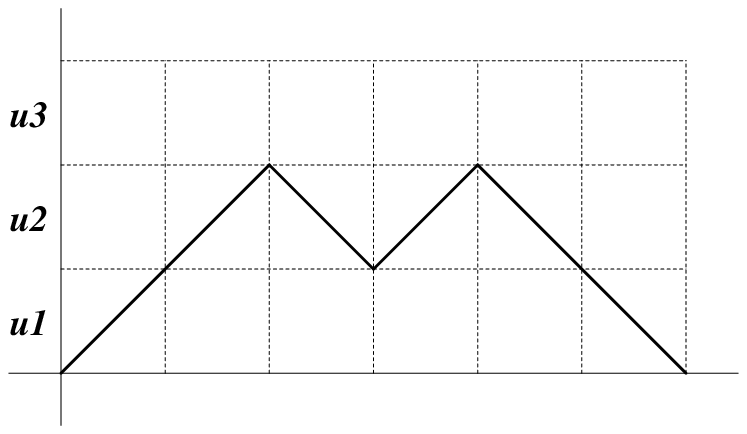} & \includegraphics[scale=0.65]{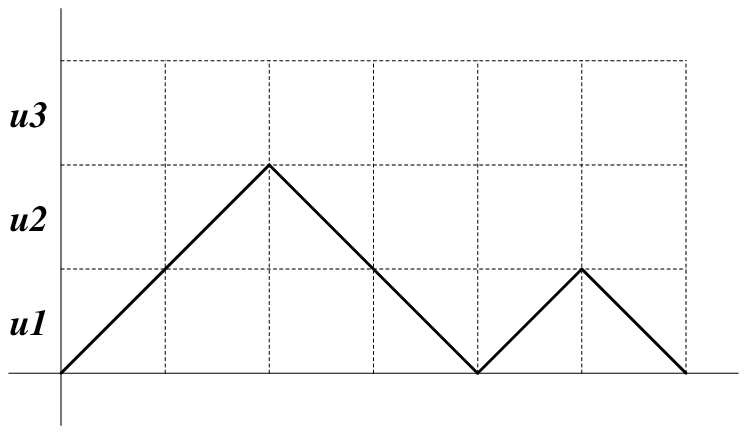} \\[1mm]
\includegraphics[scale=0.65]{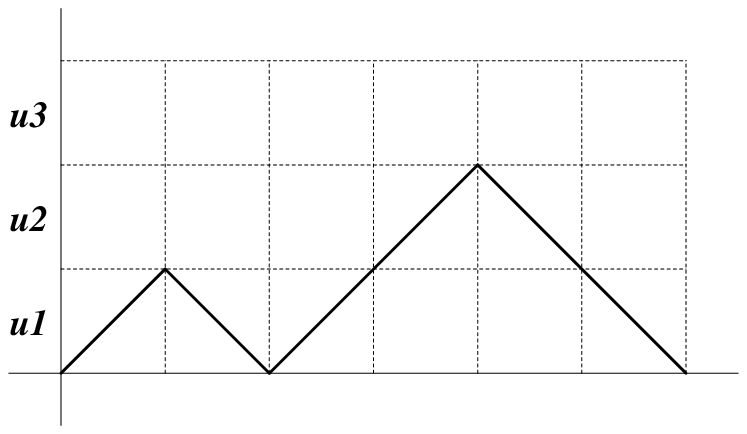} &  \includegraphics[scale=0.65]{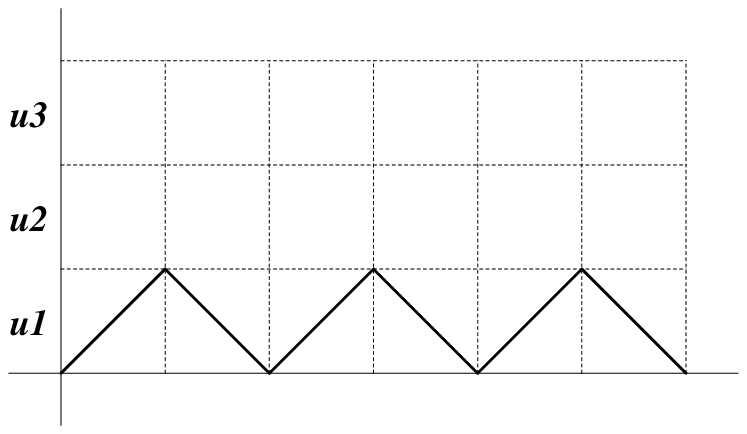} &   
\end{tabular} 
\]
\caption{All Dyck paths of size $r=3$.}
\label{fig2}
\end{figure}

\begin{figure}[hb]
\[
\begin{tabular}{ccc}
\includegraphics[scale=0.65]{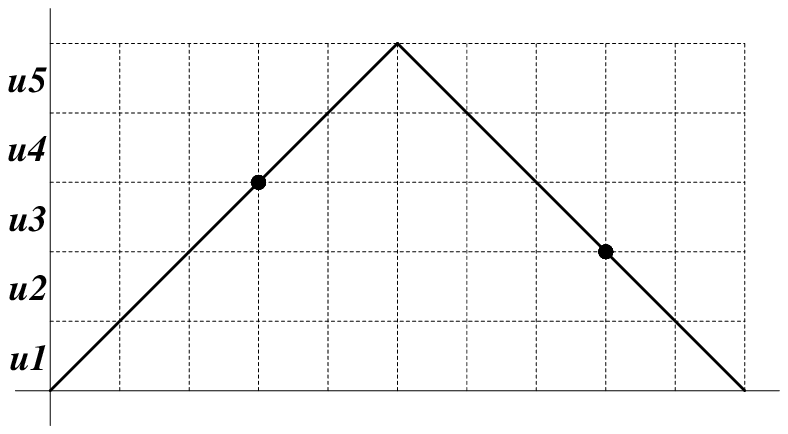} &  \includegraphics[scale=0.65]{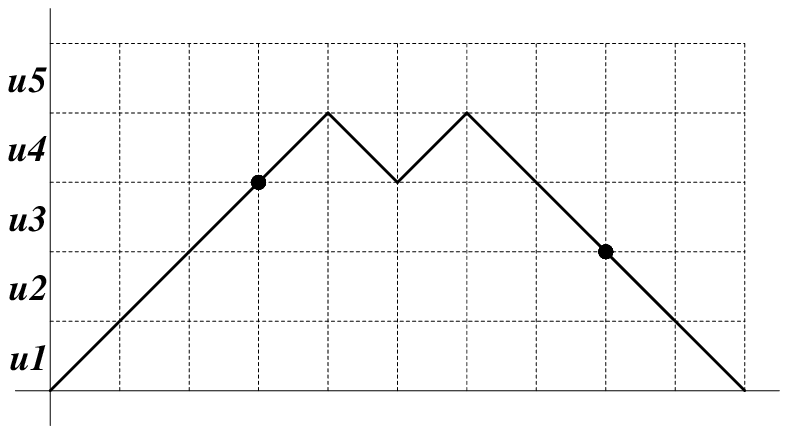} & \includegraphics[scale=0.65]{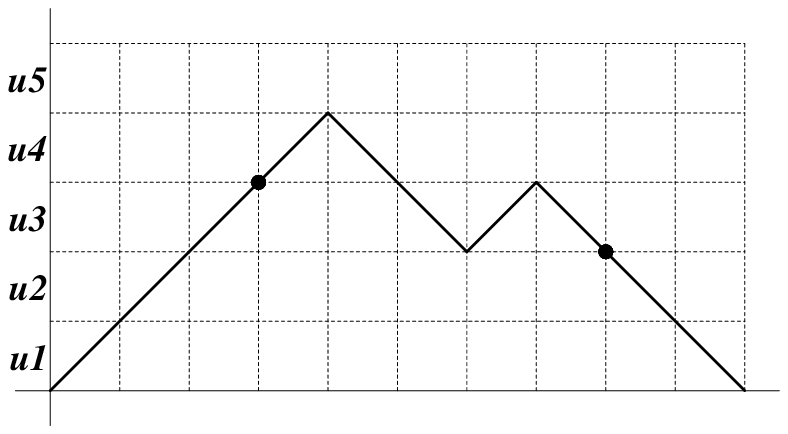} \\[1mm]
\includegraphics[scale=0.65]{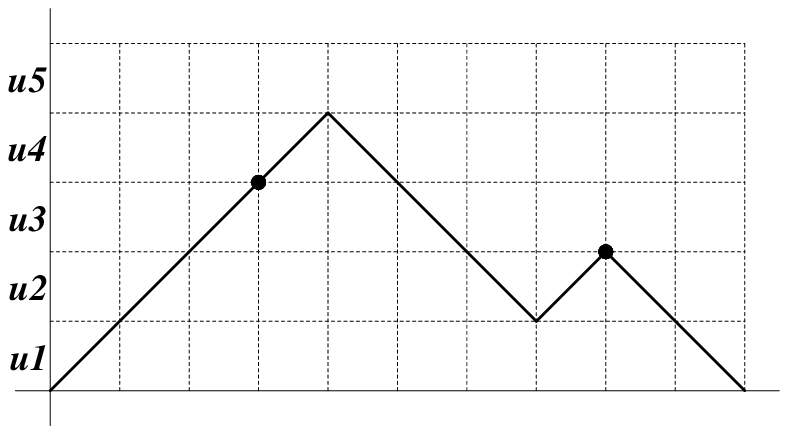} &  \includegraphics[scale=0.65]{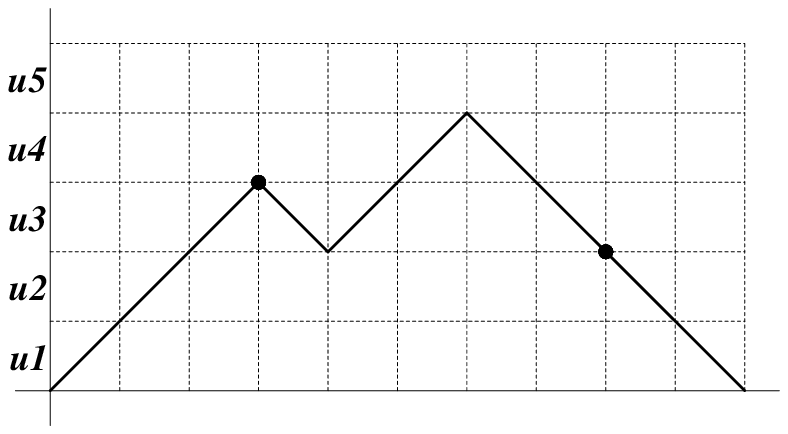} & \includegraphics[scale=0.65]{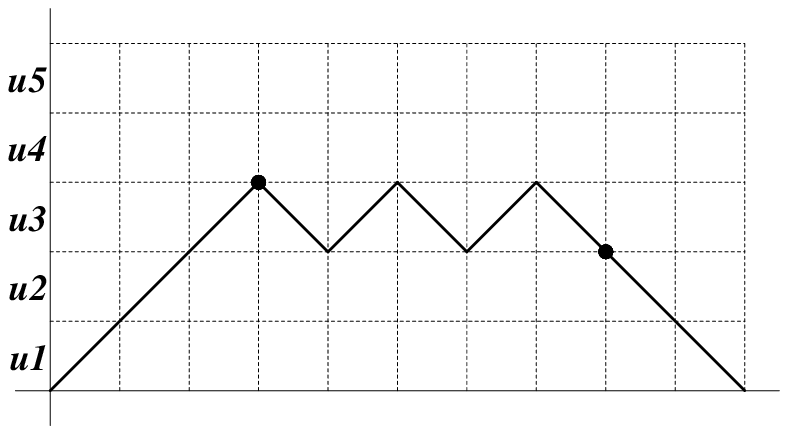} \\[1mm]
\includegraphics[scale=0.65]{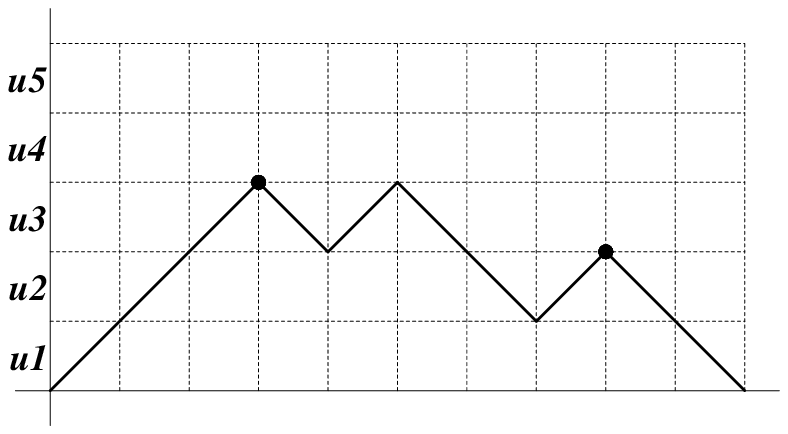} &  \includegraphics[scale=0.65]{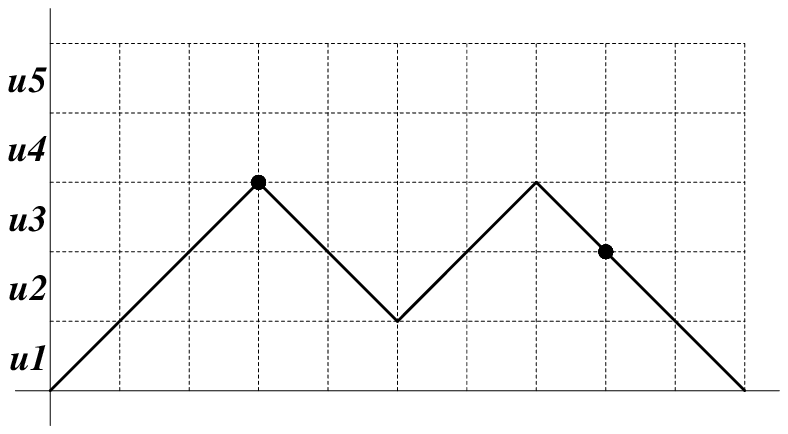} & \includegraphics[scale=0.65]{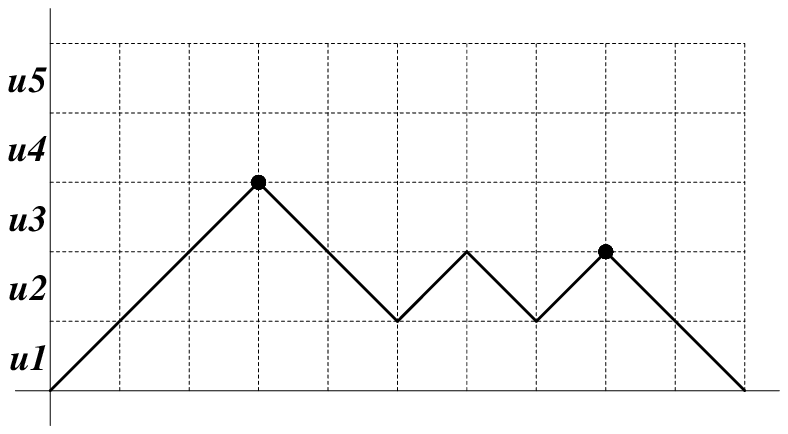} \\[1mm]
\includegraphics[scale=0.65]{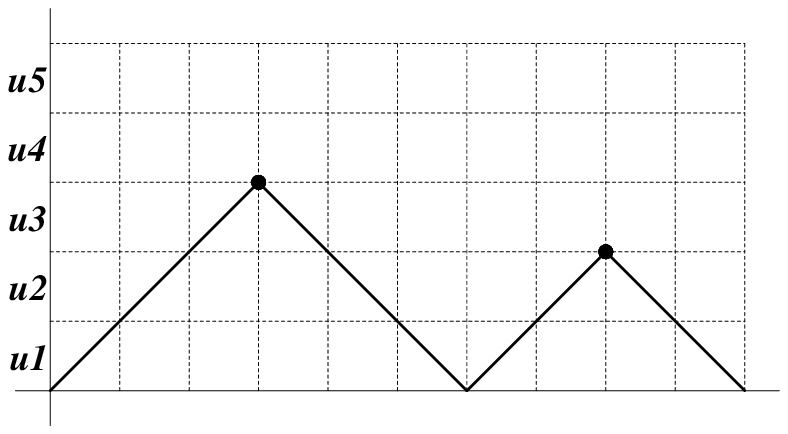} &   & 
\end{tabular} 
\]
\caption{All Dyck paths of size $r=5$ with $a=3$ and $b=2$, i.e.\ starting with at least three up steps and
ending with at least two down steps.}
\label{fig3}
\end{figure}

\end{document}